\documentclass[12pt]{iopart}
\usepackage{iopams}  
\usepackage{graphicx}
\usepackage[breaklinks=true,colorlinks=true,linkcolor=blue,urlcolor=blue,citecolor=blue]{hyperref}
\usepackage{color}
\usepackage{cite}
\usepackage{epstopdf}
\usepackage{eqnarray}

\begin{document}

\title[Quality of a densely packed multiplexed set of orthogonal modes]{Multiplexing 200 modes on a single digital hologram}

\author{Carmelo Rosales-Guzm\'an$^{1,*}$, Nkosiphile Bhebhe$^1$, Nyiku Mahonisi$^1$ and Andrew Forbes$^1$}

\address{$^1$School of Physics, University of the Witwatersrand, Johannesburg 2050, South Africa}
\ead{carmelo.rosalesguzman@wits.ac.za}
\begin{abstract}
The on-demand tailoring of light's spatial shape is of great relevance in a wide variety of research areas. Computer-controlled devices, such as Spatial Light Modulators (SLMs) or Digital Micromirror Devices (DMDs), offer a very accurate, flexible and fast holographic means to this end. Remarkably, digital holography affords the simultaneous generation of multiple beams (multiplexing), a tool with numerous applications in many fields. Here, we provide a self-contained tutorial on light beam multiplexing. Through the use of several examples, the readers will be guided step by step in the process of light beam shaping and multiplexing. Additionally, on the multiplexing capabilities of SLMs to provide a quantitative analysis on the maximum number of beams that can be multiplexed on a single SLM, showing approximately 200 modes on a single hologram.
\end{abstract}

\section{Introduction}
Pioneering work in laser beam shaping can be traced back to the 1980s but its full development was only attained in the following decades \cite{Dickey2014}. Over this period, different techniques were elucidated to reshape amplitude and phase, either internal or external to resonator cavities. In these techniques the use of amplitude and phase as well as refractive, adaptive or diffractive optical elements was very common. Along this line, holographic methods had emerge as one of the most powerful techniques, prompted by the development of modern computer-controlled devices, such as Spatial Light Modulators (SLMs) and Digital Micromirror Devices (DMDs) \cite{Forbes,Ngcobo2013,Forbes2016,OptExpress}. SLMs are pixelated screens formed by several hundreds of thousands of cells filled with liquid crystal molecules. This technology can be implemented either with transparent or reflective liquid crystals on silicon, LCDs and LCoS respectively. The molecular alignment within the pixels dictates the type of modulation the SLM imparts to the light beam: phase-only, amplitude-only or both amplitude and phase. The device is controlled via a calibrated voltage that rotates the liquid crystal molecules by a predefined angle, resulting in a form of birefringent material for a particular orientation of incoming polarized light. The voltage is regulated pixel by pixel using digital holograms, gray scale images with 255 tones of gray. A local phase shift, induced by the rotation of the molecules, takes place at each pixel of the SLM, resulting in the reshaping of an input field. 

Along with a very high accuracy in the generated light fields  \cite{Matsumoto2008,Ando2009}, digital holographic devices are very flexible, a feature that has afforded, the development of friendly-user multitouch screens for optical tweezers \cite{McDonald2013,Grieve2009,Bowman2011}. This flexibility allows also for the rapid reconfiguration of the digital hologram \cite{Thalhammer2013,Radwell2014}, to change the shape and/or position of the output field, a feature that has shown advantages in optical tweezers \cite{Curtis2002,Grier2002}, particle tracking \cite{Rosales2014OL,Rosales2014OE} and the characterization of complex light fields \cite{Perez-Vizcaino2012,Dudley2014,Schulze2015,Webster2017,Ndagano2016,Cox2016} among many others. Interestingly, SLMs afford the simultaneous generation of multiple light beams, a feature that also enables a great variety of applications \cite{Mas2011,Davis1989,Maurer2010,Maurer2011,Trichili2016}. Multiplexing is granted due to the superposition principle in optics which, by virtue of digital holography, allows the addition (multiplexing) of several holograms written into a single one. Each hologram of the superposition can be programmed to generate specific beam shapes at particular locations in the observation plane. The simultaneous generation of multiple beams has being widely exploited, for example, in the generation of multiple optical traps \cite{DGrier2003,Cismar2010,Eriksen2002} or the implementation of mode division multiplexing for optical communication \cite{Trichili2016,Wang2}.

Here we provide a self-contained tutorial aimed at the simultaneous generation of approximately 200 spatial modes encoded on a single hologram. To this end, we first introduce the Laguerre-Gaussian (LG) and Hermite-Gaussian (HG) infinite basis sets (Sec. \ref{SML}), which we use as our encoding modes. Afterwards, we explain in detail two commonly used techniques for beam shaping: phase-only and complex amplitude modulation, followed by an explanation on the generation of the digital holograms (Sec. \ref{ExpGen}) for beam multiplexing. In Sec. (\ref{ExpRes}) we describe the experimental implementation of this technique along with a quantitative analysis about the maximum number of modes that can be multiplexed on SLMs. The quality of the generated beams is determined through 2D image correlation and we show approximately 200 modes on a single hologram. 
    
\section{Laguerre-Gaussian and Hermite Gaussian modes.}
\label{SML}
Any electromagnetic field traveling in free-space obeys the wave equation. Well-known solutions to the wave equation in the paraxial regime can be given in cylindrical and Cartesian coordinates. The former is given in terms of the generalized Laguerre polynomials $L_p^\ell(x)$ modulated by a Gaussian envelope, which forms an infinite orthogonal set of solutions, known as the Laguerre-Gaussian, $LG_p^\ell$, modes \cite{Siegman}:

\begin{eqnarray}
\hspace{-18mm}
\nonumber
LG_p^\ell(\rho,\varphi,z)= \sqrt{\frac{2p!}{\pi(\ell+p)!}}\left[\frac{\sqrt{2}\rho}{\omega(z)}\right]^{\ell}L_p^\ell\left[\frac{2\rho^2}{\omega^2(z)}\right]\frac{\exp[i(2p+\ell+1)\zeta(z)]}{\omega(z)}\\\vspace{4mm} \exp\left[-\frac{\rho^2}{\omega^2(z)}\right]\exp\left[\frac{-ik\rho^2}{2R(z)}\right] \exp[-i\ell\varphi],
\label{LG}
\end{eqnarray}

\begin{figure}[tb]
\includegraphics[width=1\textwidth]{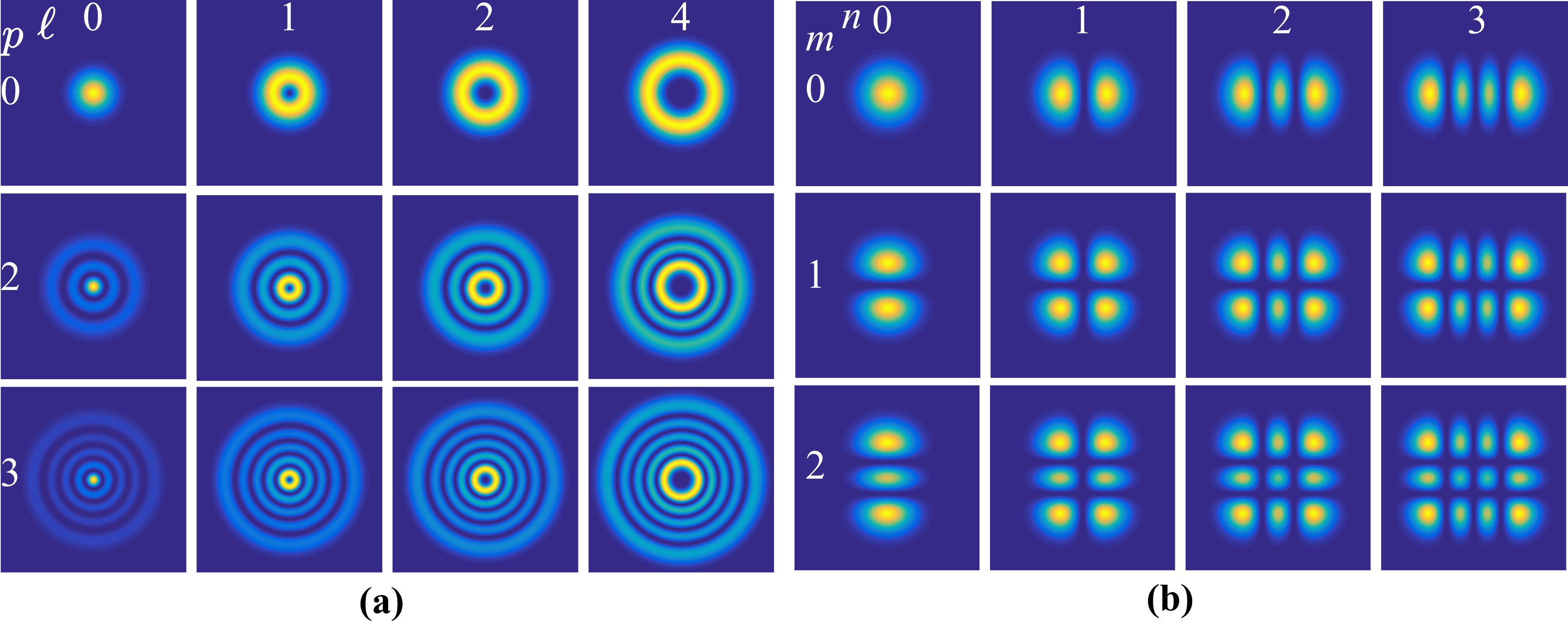}
\caption{ \textbf{Theoretical 2D instensity profiles of $LG_{p}^\ell$ and $HG_{nm}$ modes}. Intensity distribution of  (a) LG modes with $l=[0, 1, 2, 3,4]$, $p=[0, 2, 3]$ and (b) HG modes with $n=[0, 1, 2, 3]$ $m=[0, 1, 2]$.}
\label{HGLGmodes}
\end{figure}

\vspace{5mm}
\noindent
where, $(\rho,\varphi,z)$ is the position vector of the cylindrical coordinates, $\ell$ is an integer number that accounts for the number of times the phase wraps around the optical axis, $p$ is a positive integer that accounts for the number of maxima ($p+1$) along the transverse direction. Each solution represents a paraboloidal wave with radius of curvature $R(z)=z[1+(z_R/z)^2]$, beam waist $\omega_0$ and Rayleigh range $z_R$; $\omega(z)=\omega_0\sqrt{1+(z/z_R)^2}$ is the beam width as function of $z$ and $\zeta(z)=\arctan(z/z_R)$ is the Gouy phase. Figure \ref{HGLGmodes}(a) shows the intensity distributions of the first few $LG_p^\ell(x)$ modes with combinations of $p=[0, 1, 3]$ and $\ell=[0, 1, 2, 4]$.

In Cartesian coordinates, the paraxial wave equation allows as solutions the Hermite-Gaussian modes, $HG_{nm}$, a product of Hermite polynomials $H(x)$ and a Gaussian envelope, with $n$ and $m$ positive integers: 
\begin{eqnarray}
\nonumber
\hspace{-18mm}
HG_{nm}(x,y,z)=\frac{1}{\omega(z)}\sqrt{\frac{2^{-(n+m-1)}}{\pi n!m!}}\exp\left[i(n+m+1)\zeta(z)\right]H_n\left[\frac{\sqrt{2}x}{\omega(z)}\right]H_m\left[\frac{\sqrt{2}y}{\omega(z)}\right]\\\vspace{4mm}\exp\left[-\frac{x^2+y^2}{\omega^2(z)}\right]\exp\left[\frac{-ik(x^2+y^2)}{2R(z)}\right]\exp[-ikz].
\label{HermiteGaussian}
\end{eqnarray}
\noindent
The parameters $\omega$, $\omega_0$, $z_R$, $R(z)$ and $\zeta(z)$ are the same as for the $LG_{p}^\ell$ modes. The positive integer subindexes $n$ and $m$ are related to the number of bright lobes of the transverse intensity profile along $x$ and $y$ respectively. Figure \ref{HGLGmodes} (b) shows the intensity profile of the modes described by combinations of $n=[0, 1, 2, 3]$ and $m=[0, 1, 2]$.

\section{Experimental generation and multiplexing of modes}
\label{ExpGen}

Shaping of light fields in both amplitude and phase, requires in general a transfer function capable of modulating both, amplitude and phase. Since most SLMs allow only the direct control of the phase, several techniques have being proposed to indirectly modulate the amplitude \cite{Ohtake2007,Matsumoto2008,Arrizon,Davis2003,Putten2008,Clark2016,Ando2009,Aguirre2015}. In this section we will briefly explain two of these techniques, phase-only and complex amplitude modulation. Here we will also explain the general basics through which multiple beams can be generated using a single hologram.

\subsection{Mode generation and multiplexing using phase holograms}

\begin{figure}[h!]
\includegraphics[width=\textwidth]{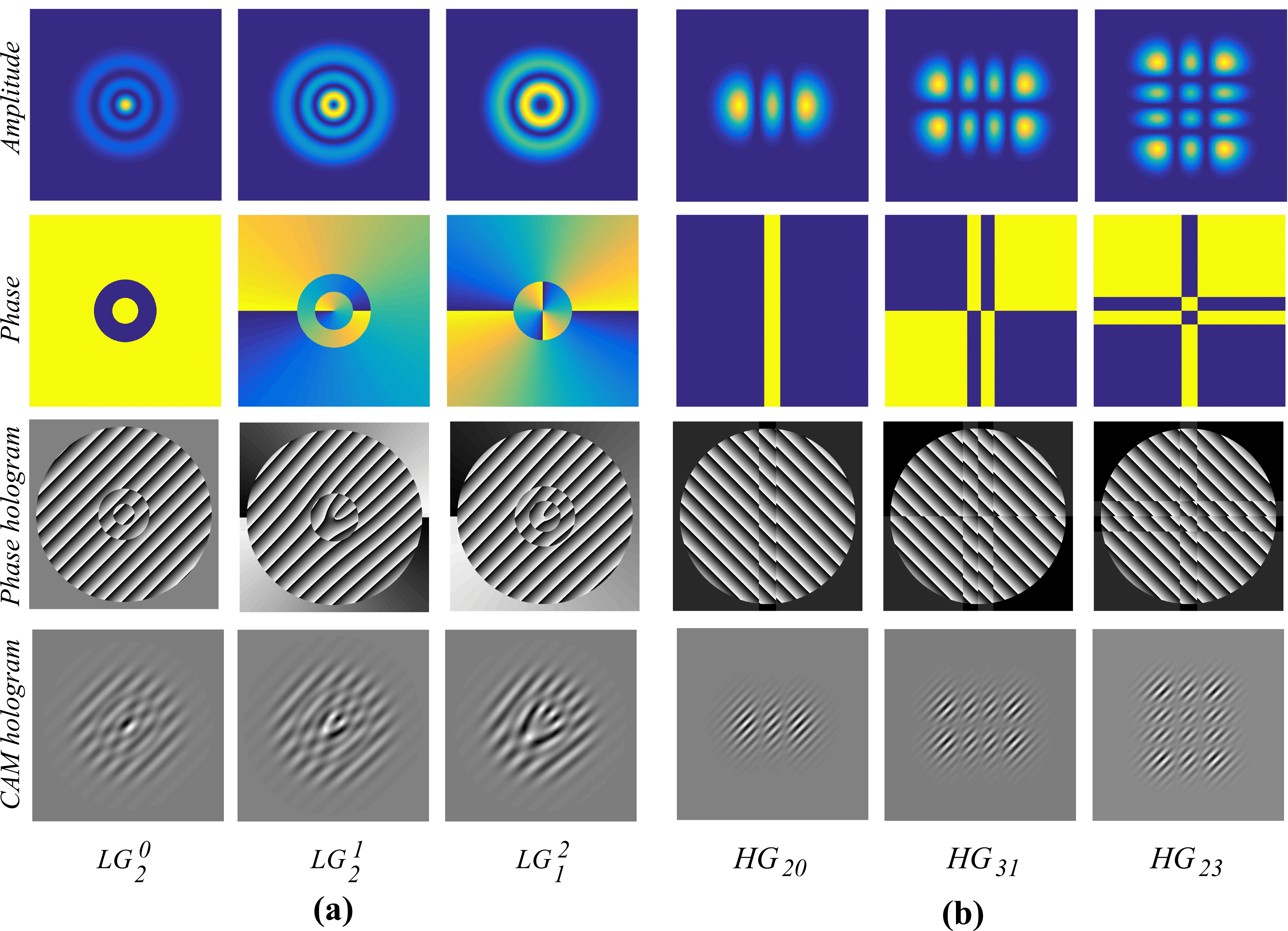}
\caption{ \textbf{Digital holograms, phase and amplitude for $LG_{p}^\ell$ and $HG_{nm}$ modes.} The first row shows the amplitude while the corresponding phase profiles are given in the second row. Phase only and complex-amplitude modulation holograms shown in the third and fourth row, respectively were encoded on the SLM to generate (a) $LG_{p}^\ell$ and (b) $HG_{mn}$ modes.}
\label{Holograms}
\end{figure}

In order to generate a phase-only hologram capable to reproduce the amplitude changes in $LG_p^\ell$ modes, it is sufficient, in a first approximation, to encode the changes of sign given by the Laguerre polynomial $L_p^\ell$ in the radial direction. This, in combination with azimuthal phase variation, generates the transfer function\cite{Matsumoto2008},

\begin{equation}
t(\rho,\varphi)=\exp\left\{i\left[\ell\varphi+\pi\Theta(\rho)\right]\right\},
\end{equation}
\noindent
where, $\Theta(\rho)$ is the Heaviside function that takes into account the sign variations of the Laguerre-Gauss function and is given by,

\begin{equation}
\Theta(\rho)=\Theta\left[L_p^{|\ell|}\left(\frac{2\rho^2}{\omega_0^2}\right)\right].
\end{equation}
\noindent
The above transfer function can be encoded on the SLM along with a linear phase grating, to separate the different diffraction orders. The transfer function of a linear phase grating is given by,
\begin{equation}
t_g(x,y)=\exp\left[i2\pi\left(\frac{x}{x_0}+\frac{y}{y_0}\right)\right],
\end{equation}
where $x_0$ and $y_0$ are the spatial periods of the grating along the horizontal and vertical direction respectively. A lens of focal length $f$ is placed at a distance $f$ from the SLM to separate the diffraction orders in the focal plane of the lens. The final coordinates $x_f$ and $y_f$ of the first diffraction order are related to the grating period and the focal length $f$ by the relations $x_f=\lambda f/x_0$ and $y_f=\lambda f/y_0$, respectively, where $\lambda$ is the wavelength of the incident field. Hence, the transfer function in terms of $x_f$ and $y_f$ takes the form,
\begin{equation}
t_g(x,y)=\exp\left[i2\pi\left(x\frac{x_f}{\lambda f}+y \frac{y_f}{\lambda f}\right)\right].
\end{equation}
Hence, the hologram to generate the $LG_p^\ell$ mode combined with the diffraction grating takes the final form, 
\begin{equation}
\Phi=\bmod\left\{\ell\varphi+\pi\Theta\left[L_p^{|\ell|}\left(\frac{2\rho^2}{\omega_0^2}\right)\right]+2\pi\left(x\frac{ x_f}{\lambda f}+y\frac{y_f}{\lambda f}\right),2\pi\right\}.
\end{equation}
\noindent
In the above equation, $\bmod[x]$ represents the modulus function that wraps the phase around $2\pi$. A similar approach can be followed to generate $HG_{nm}$ beams \cite{Aguirre2015}. Figure \ref{Holograms} shows the amplitude, phase and holograms encoded on the SLM for $LG_p^\ell$ \ref{Holograms}(a) and $HG_{nm}$ \ref{Holograms}(b) modes.

\begin{figure}[tb]
\includegraphics[width=\textwidth]{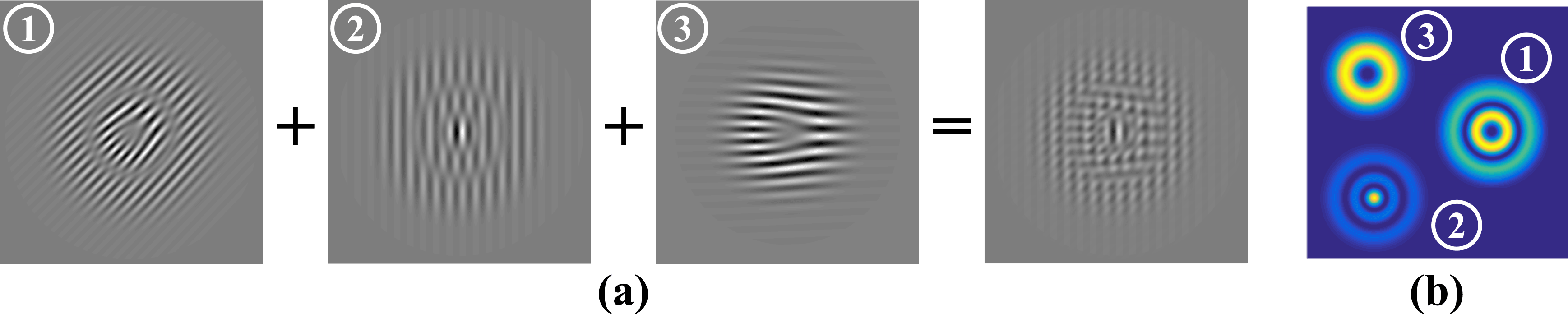}
\caption{\textbf{Multiplexing concept}. (a) A superposition of holograms with unique carrier frequencies (grating) forming a single hologram encoded on to the SLM. The unique carrier frequencies correspond to a specific spatial position, on the Fourier plane, of the encoded modes as shown in (b).}
\label{Multip}
\end{figure}

To generate multiple beams simultaneously, we combined on a single hologram a superposition of multiple holograms, each with a unique spatial carrier frequency, as illustrated in Fig. \ref{Multip}(a). Clearly, each spatial frequency $(1/x_0, 1/y_0)$ directs the generated beam to a particular location in the Fourier plane, as illustrated in Fig \ref{Multip}(b). Careful selection of the spatial frequencies allows a proper distribution of the multiplexed beams to avoid any overlapping. The final hologram $\Phi_M$ for M multiplexed beams takes the form,
\begin{equation}
\Phi_M=\sum_{j=1}^M\bmod\left[\ell_j\varphi+\pi\Theta_j(\rho)+2\pi\left(x\frac{x_f^j}{\lambda f}+y\frac{y_f^j}{\lambda f}\right),2\pi\right].
\end{equation}
The specific locations of each mode can be selected by properly choosing $\ell_j$, $\Theta_j(\rho)$ and ($x_f^j$, $y_f^j$).
\subsection{Mode generation and multiplexing using complex amplitude modulation}

Complex amplitude modulation can be achieved in many ways, in this tutorial we will only focus on the method proposed by Arrizon {\it et al.} \cite{Arrizon} that produces the highest quality mode compared to other methods \cite{Ohtake2007,Clark2016}. To start with, lets consider a field of amplitude $A({\bf r})$ and phase $\phi({\bf r})$ of the form,
\begin{equation}
\label{CAM}
U({\bf r})=A({\bf r})\exp[i\phi({\bf r})]
\end{equation}
\noindent 
with A({\bf r}) and $\phi({\bf r})$ in the intervals [0,1] and $[-\pi,\pi]$, respectively. One way to indirectly modulate the amplitude from phase modulation can be done by writing Eq. \ref{CAM} as
\begin{equation}
\label{h}
h({\bf r})=\exp\{i\Phi[A({\bf r}),\phi({\bf r})]\}
\end{equation}
\noindent
Everything reduces to finding the function $\Phi[A({\bf r}),\phi({\bf r})]$. This can be done by expanding Eq. \ref{h} as a Fourier series of the form
\begin{equation}
\label{h_xy}
h({\bf r})=\sum_{q=-\infty}^\infty c_q^A\exp[iq\phi({\bf r})],
\end{equation}
where,
\begin{equation}
c_q^A=\frac{1}{2\pi}\int_{-\pi}^\pi\exp\{i\Phi[A({\bf r}),\phi({\bf r})]\}\exp[-iq\phi({\bf r})]d\phi.
\end{equation}
The field $U({\bf r})$ (Eq. \ref{CAM}), can be recovered from the first term of the above equation provided,
\begin{equation}
\label{c1q}
c_1^A=\frac{1}{2\pi}\int_{-\pi}^\pi\exp\{i\Phi[A({\bf r}),\phi({\bf r})]\}\exp[-i\phi({\bf r})]d\phi=A({\bf r})a,
\end{equation}
with $a$ a positive constant. Equation \ref{c1q} can be fulfilled if,
\begin{equation}
\label{sine}
\int_{-\pi}^\pi\sin\{i\Phi[A({\bf r}),\phi({\bf r})]\}d\phi=0,
\end{equation}
\begin{equation}
\label{cosine}
\int_{-\pi}^\pi\cos\{i\Phi[A({\bf r}),\phi({\bf r})]\}d\phi=2\pi a,
\end{equation}
which provides an appropriate basis to determine $\Phi[A({\bf r}),\phi({\bf r})]$. In Eq. \ref{cosine} the value of the constant $a$ is in the interval [0,1], which limits the efficiency of the generated hologram. While different solutions to Eqs. \ref{sine} and \ref{cosine} can be found, here we will only focus on a solution with odd symmetry, namely, 
\begin{equation}
\label{eq14}
\Phi[A({\bf r}),\phi({\bf r})]=f[A({\bf r})]\sin[\phi({\bf r})],
\end{equation}
which corresponds to type 3 hologram in reference \cite{Arrizon}, according to which $h({\bf r})=\exp\{if[A({\bf r})]\sin[\phi({\bf r})]\}$ can be expanded, using the Jacobi-Anger identity as,
\begin{equation}
\label{eq15}
\exp\{if[A({\bf r})]\sin[\phi({\bf r})]\}=\sum_{s=-\infty}^\infty J_s\{f[A({\bf r})]\}\exp[is\phi({\bf r})],
\end{equation}
where, $s$ is an integer and $J_s$ are the Bessel function. Direct comparison of Eqs. \ref{eq15} and \ref{h_xy} yields,
\begin{equation}
c_q^A({\bf r})=J_q\{f[A({\bf r})]\}.
\end{equation}
Equation \ref{c1q} yields the final expression
\begin{equation}
\label{eq17}
c_1^A({\bf r})=J_1\{f[A({\bf r})]\}=A({\bf r})a,
\end{equation}
where $J_1(x)$ is the first order Bessel function. Finally, $f[A({\bf r})]$ can be obtained by numerical inversion of Eq. \ref{eq17} as,
\begin{equation}
\label{Bessel}
f[A({\bf r})]=J_1^{-1}[A({\bf r})a]
\end{equation}
The inversion of the $J_1(x)$ is guaranteed in the interval $[0,x_m]$, being $x_m=1.84$ the value where $J_1(x)$ reaches its first maximum $J_1(x_m)=0.5819$. This restriction limits the values of $\phi[A({\bf r}),\phi({\bf r})]$ to the reduced domain $[-0.586\pi,0.586\pi]$, which implies that the maximum phase shift that can be obtained with this method is $\Delta\phi=1.17\pi$.

To generate the required holograms to transform an input light field into $LG_p^\ell$ or $HG_{nm}$, we can consider the SLM to be located at $z=0$ and simplify Eqs. \ref{LG} and \ref{HermiteGaussian} as,
\begin{equation}
\hspace{-18mm}
LG_p^\ell(\rho,\varphi,z)= \frac{1}{\omega_0}\sqrt{\frac{2p!}{\pi(\ell+p)!}}\left[\frac{\sqrt{2}\rho}{\omega_0}\right]^{\ell}L_p^\ell\left[\frac{2\rho^2}{\omega_0^2}\right]\exp\left[-\frac{\rho^2}{\omega_0^2}\right]\exp[-i\ell\varphi]
\label{LG0}
\end{equation}
and
\begin{equation}
\hspace{-18mm}
HG_{nm}(x,y,z)=\frac{1}{\omega_0}\sqrt{\frac{2^{-(n+m-1)}}{\pi n!m!}}H_n\left[\frac{\sqrt{2}x}{\omega_0}\right]H_m\left[\frac{\sqrt{2}y}{\omega_0}\right]\exp\left[-\frac{x^2+y^2}{\omega_0^2}\right],
\label{HGauss}
\end{equation}
respectively. The phase $\phi({\bf r})$ and amplitude $A({\bf r})$ can be easily obtained from the above equations and inserted into Eq. \ref{Bessel} and finally in Eq. \ref{eq14} to obtain the desired hologram, which will take the form
\begin{equation}
\hspace{-18mm}
\Phi=\bmod\left\{J_1^{-1}[0.5819A({\bf r})]\sin\left[\phi({\bf r})+1.17\pi\left(x\frac{x_f}{\lambda f}+y\frac{y_f}{\lambda f}\right)\right],\hspace{1mm} 1.17\pi\right\}.
\end{equation}
Figure \ref{Holograms} (fourth row) shows examples of the holograms obtained as described above to generate $LG_p^\ell$ and $HG_{nm}$ modes. Multiplexing of $LG_P^\ell$ and $HG_{mn}$ can be achieved by encoding a hologram of the form,
\begin{equation}
\hspace{-18mm}
\Phi_M=\sum_{i=1}^M\bmod\left\{J_1^{-1}[0.5819A_i({\bf r})]\sin\left[\phi_i({\bf r})+1.17\pi\left(x\frac{ x_f^i}{\lambda f}+y\frac{ y_f^i}{\lambda f}\right)\right],\hspace{1mm} 1.17\pi\right\},
\end{equation}
where, $A({\bf r})_i$ and $\phi({\bf r})_i$ represents the amplitude and phase of the encoded mode, respectively, $x_f^i$ and $y_f^i$ are the are the final positions of each mode in the vertical and horizontal direction, respectively. Figure \ref{Multip} shows multiplexing of three $LG_p^\ell$ modes $LG_1^1$, $LG_2^0$ and $LG_0^2$ with different ($x_0^i$, $y_0^i$) values.

\section{Experimental Implementation of mode multiplexing}
\label{ExpRes}

\subsection{Experimental setup}
\begin{figure}[b]
\includegraphics[width=\textwidth]{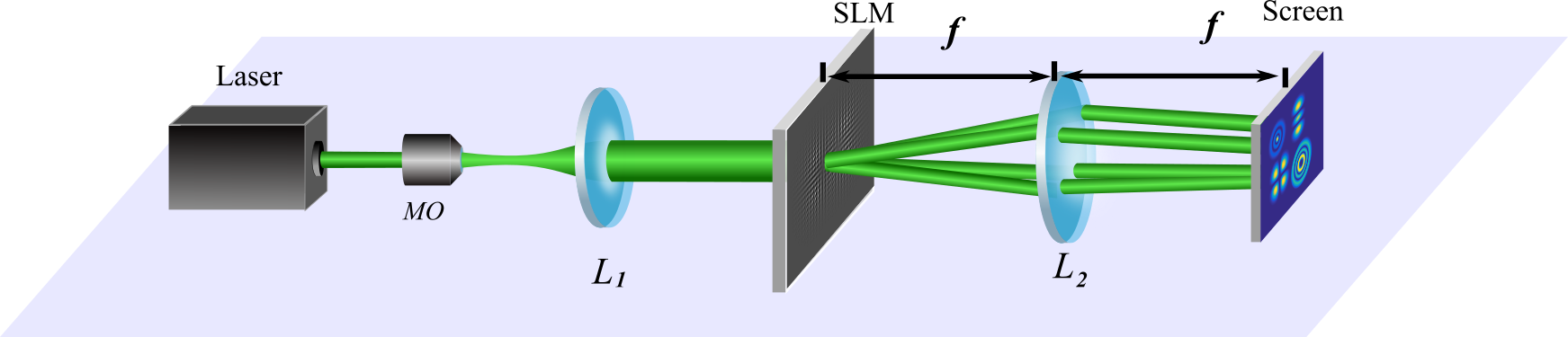}
\caption{\textbf{Schematic of the mode multiplexing setup}.The laser beam (532 nm) is expanded using a 10$\times$ microscope objective (MO) and a collimating lens $L_1$ on to a spatial light modulator (SLM). A hologram of multiplexed $LG_{p}^\ell$ and $HG_{mn}$ modes is addressed on the SLM. The desired modes are generated at focal plane of the Fourier lens $L_2$ with unique spatial positions as shown on the screen.}
\label{Exp}
\end{figure}

A schematic of the multiplexing setup is shown in Fig. \ref{Exp}. The laser source, a 532 nm linearly polarized Verdi G Coherent laser was expanded using a telescope constituting a $\times$10 microscope objective and a lens $L_1$ of focal length $f = $ $100$ $mm$ approximating a plane wave. The expanded beam impinged onto a reflective Holoeye Pluto spatial light modulator (SLM) of 1920 $\times$ 1080 pixel dimensions with a pixel pitch of $8$  $\mu m$. The SLM screen was split into three sections which enabled independent hologram addressing on each third. A multiplexed hologram to generate $LG_p^\ell$, $HG_{nm}$ or a combination of modes was encoded on each third of the SLM. Each mode was encoded with a unique carrier frequency, as explained in section \ref{ExpGen}, to separate all modes in the far field. The far field was obtained by placing a lens ($L_2$) of focal length $f = $ $250$ $mm$ a distance $f$ from the SLM and placing the detection plane at a distance $f$ from $L_2$. All the multiplexed modes were generated in the $1^{st}$ diffraction order which was separated from the rest using a spatial filter. The mode quality of the generated modes was tracked using 2D image correlation, as will be explained in the next section.   

\subsection{Measurement of mode quality}
The quality of the experimentally generated modes was quantified by correlating their intensity profile against their theoretical counterpart, represented by images $A$ and $B$ respectively. The 2D images correlation was computed as,
\begin{equation}
C=\frac{\sum_i\sum_j(A_{ij}-\bar{A})(B_{ij}-\bar{B})}{\sqrt{\sum_i\sum_j(A_{ij}-\bar{A})^2\sum_i\sum_j(B_{ij}-\bar{B})^2}},
\end{equation}
\noindent
where $C$, the correlation coefficient, is a dimensionless parameter that measures the similarity between two images, 0 for nonidentical images and 1 for identical images. $A_{ij}$ and $B_{ij}$ are the intensity values per pixel. $\bar{A}$ and $\bar{B}$ are the mean intensity values of $A$ and $B$ respectively. 

\begin{figure}[tb]
\includegraphics[width=1\textwidth]{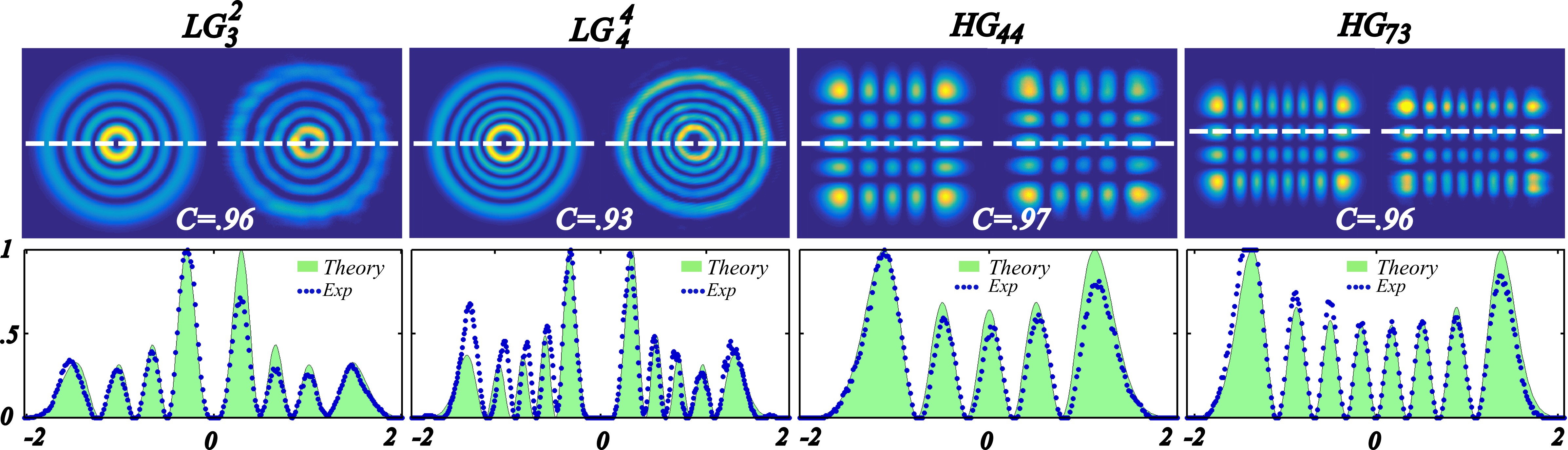}
\caption{\textbf{Mode purity measurements using correlation.}The experimental and theoretical intensity profiles for $LG_p^\ell$ and $HG_{nm}$ modes are shown in the top row together with the corresponding correlations, C. The respective transverse plots are given in the bottom row. For both $LG_p^\ell$ and $HG_{nm}$ modes, the  correlation decreases with increase in the complexity of the mode. 
}
\label{2Dcor}
\end{figure}

A straightforward correlation measurement of the generated modes provides information about their quality as a function of their structure complexity. Figure \ref{2Dcor} (top) shows the transverse intensity profile of the modes $LG_3^2$, $LG_4^4$, $HG_{44}$ and $HG_{73}$ generated through complex amplitude modulation. The corresponding correlation coefficients are $C=0.96$, $C=0.93$, $C=0.97$ and $C=0.96$. A cross-sectional plot of the mode's intensity profile, illustrating the intensity variations across the beam, is also shown in Fig. \ref{2Dcor} (bottom). It can also be observed that $HG_{nm}$ modes gives higher correlation values than $LG_p^\ell$.

\begin{figure}[tb]
\includegraphics[width=\textwidth]{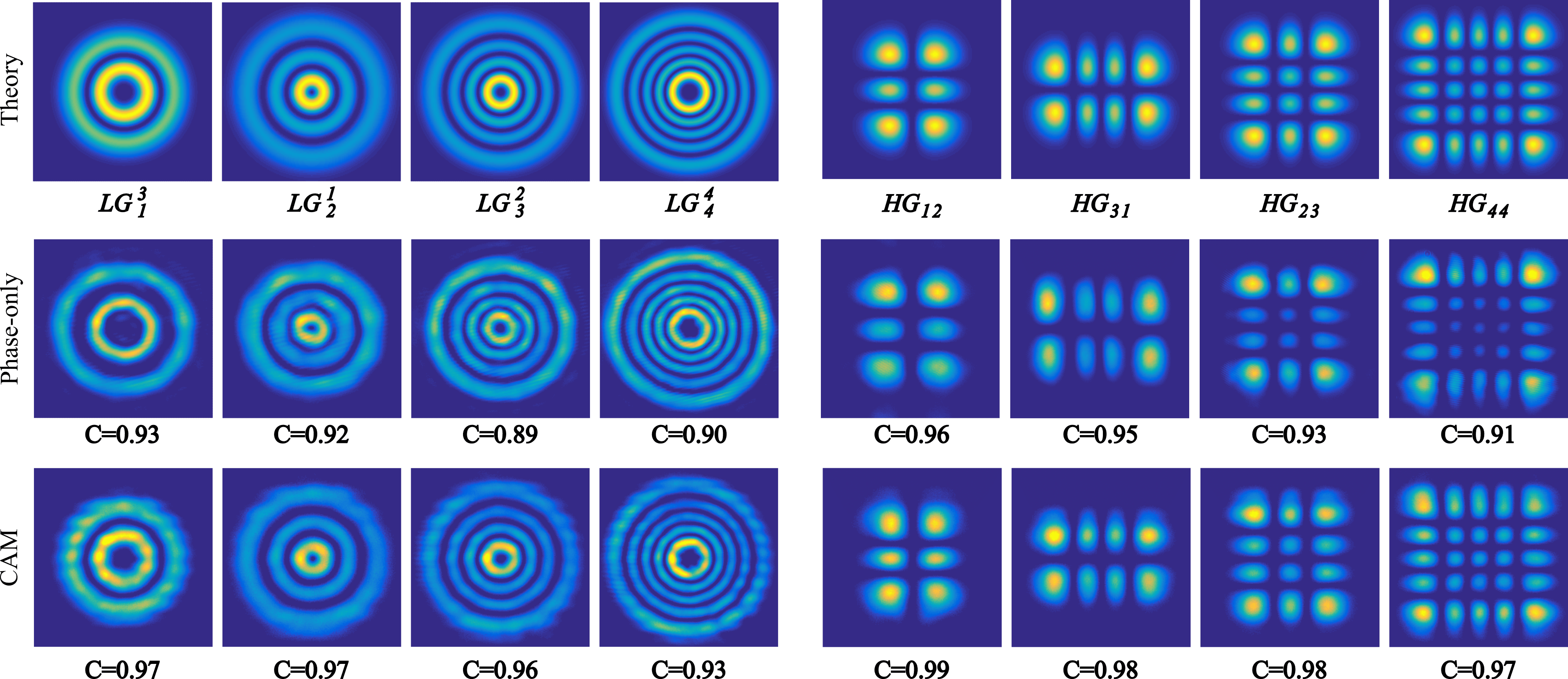}
\caption{\textbf{A comparison of mode purity for phase only and complex-amplitude modulation (CAM) generated $LG_p^\ell$ and $HG_{nm}$ modes}. The top row shows the theoretical intensity distributions with the experimental images obtained using phase only modulation given in the center row and the corresponding images obtained using CAM are in the bottom row. The modes generated using CAM present a higher correlation, C compared to the phase only modes.} 
\label{Corr}
\end{figure}

A comparison of the mode quality between the two beam shaping techniques, phase-only and complex amplitude modulation (CAM),was also performed. Figure \ref{Corr} shows the theoretical (top) intensity profile of the modes $LG_1^3$, $LG_2^1$, $LG_3^2$, $LG_4^4$, $HG_{12}$, $HG_{12}$, $HG_{31}$, $HG_{23}$ and $HG_{44}$ along with their corresponding experimental profiles generated through phase-only (Fig. \ref{Corr} center row) and CAM (Fig. \ref{Corr} bottom row). As expected, the correlation values decreases as the complexity of the modes increases. Moreover, the correlation values are always higher for modes generated through CAM.

\subsection{Beam quality of multiplexed modes}
\begin{figure}[b]
\includegraphics[width=\textwidth]{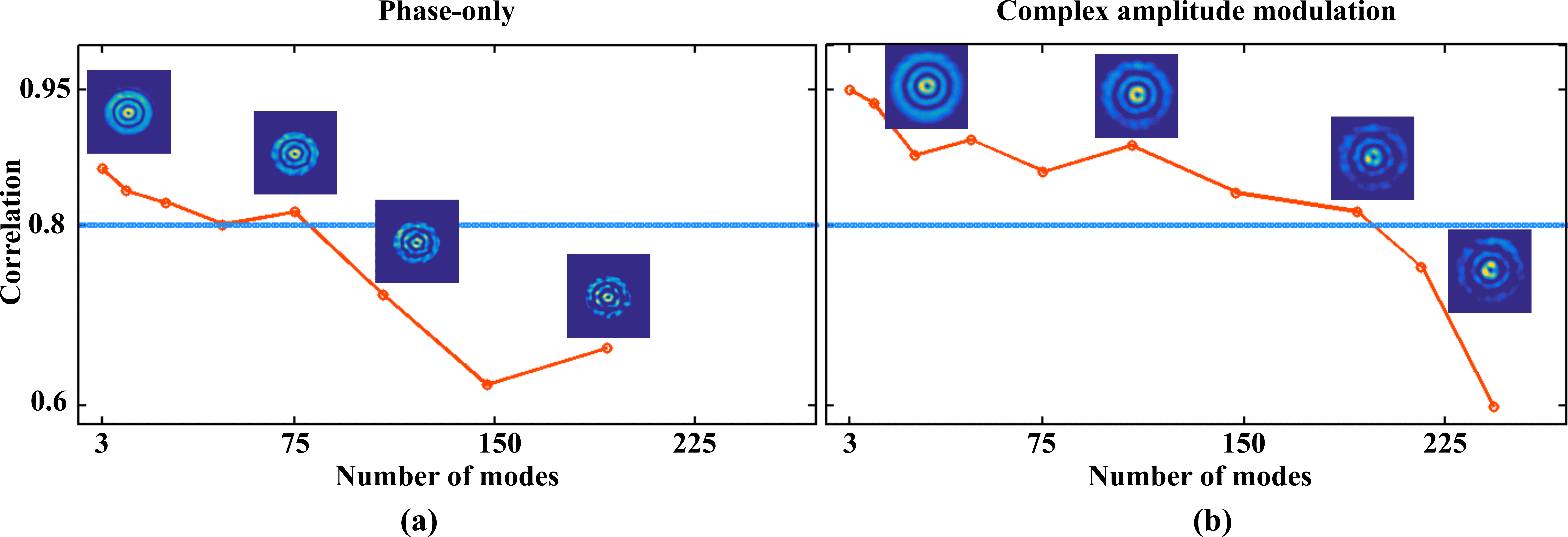}
\caption{\textbf{Correlation as a function of the number of multiplexed modes.} Plots of the correlation of an $LG_2^1$ mode with increase in the number modes obtained with (a) Phase-only and (b) CAM holograms. Only $75$ modes of good quality (i.e. above $C=0.81$) were achieved with phase-only modulation while $192$ modes can be  multiplexed with CAM keeping a correlation above the set threshold.}
\label{CorPlot}
\end{figure}

In order to quantify the maximum number of modes an SLM can support, we tracked the correlation for a specific mode for increasing number of multiplexed modes. Our threshold for a good quality mode was set at a correlation $C=0.8$, which is the maximum correlation value that a theoretical mode can achieve when compared to its own binary version, horizontal line in Fig. \ref{CorPlot}. In our experiment, we split the SLM's screen in three independent sections, each of which was addressed with its own multiplexed hologram. In this way, each third of the SLM behaves as an independent SLM and can be illuminated with the same source or with different sources and/or wavelengths \cite{Trichili2016}. 
In this way, the number of multiplexed modes increases by a factor of three. Next, we encoded arbitrary modes on each third of the SLM and tracked the correlation of one of them as we increased the number of multiplexed modes. This procedure was carried out for both multiplexing schemes and compared afterwards. Figure \ref{CorPlot} (a) shows the correlation as a function of the number of modes for phase-only holograms whereas Fig. \ref{CorPlot}(b) shows the same measurements for CAM holograms. Notice how the correlation values decays very rapidly for phase-only holograms, which according to our correlation threshold only 75 modes can be multiplexed in order to have correlation values higher than 0.8. On the contrary, for CAM, the correlation coefficient remains higher than 0.8 up to about 192 modes.

\begin{figure}[t]
\includegraphics[width=\textwidth]{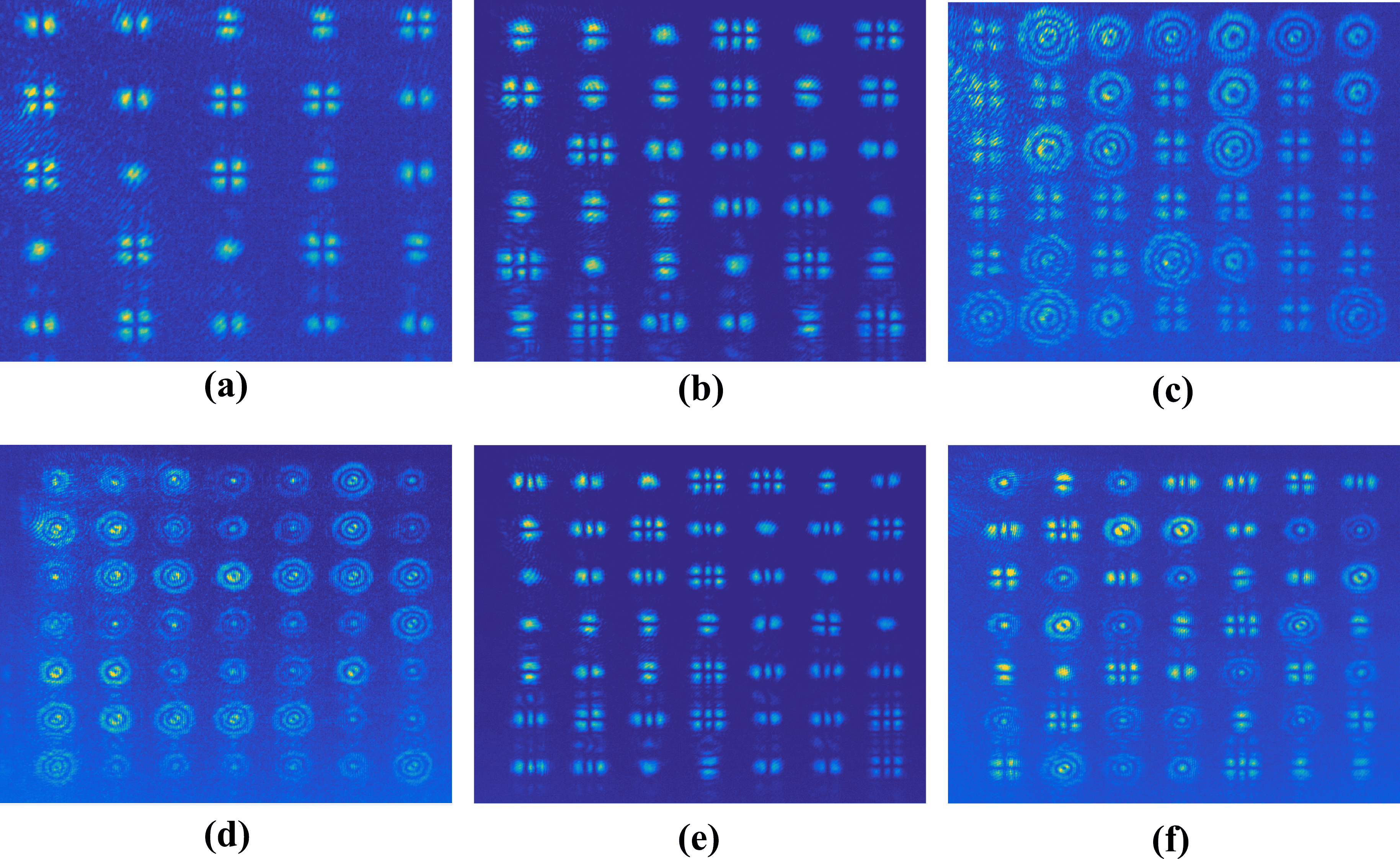}
\caption{\textbf{Spatial mode multiplexing of $LG_p^\ell$ and $HG_{nm}$ modes.} Experimental images of (a) 25, (b) 36, (c) 42 and (d)-(f) 49 multiplexed modes.  The multiplexed modes can be from the same mode set as illustrated in (a), (b), (d) and (e) i.e. $LG_p^\ell$ and $HG_{nm}$ modes encoded separately or a combination of two mode sets as shown in (c) and (f).  The images only show a maximum of $49$ multiplexed modes due to the limitations of the detector.}
\label{Multiplexed}
\end{figure}

Figure \ref{Multiplexed} shows six sets of multiplexed modes, captured with a CCD camera (BFLY-U3-13S2M-CS from  Pointgrey, $1288$ $ \times$ $964$ pixels and 3.75 $\mu m$ pixel size). By properly choosing the grating period of each multiplexed mode, we can multiplex 25, 36, 42 and 49 modes, as shown in Fig. \ref{Multiplexed}(a), (b), (c) and (d) respectively. In the figures, modes from the same set Fig. \ref{Multiplexed} (a),(b) and (d) or from both sets Fig. \ref{Multiplexed}(c) and (f) where randomly multiplexed.

\begin{figure}[tb]
\includegraphics[width=\textwidth]{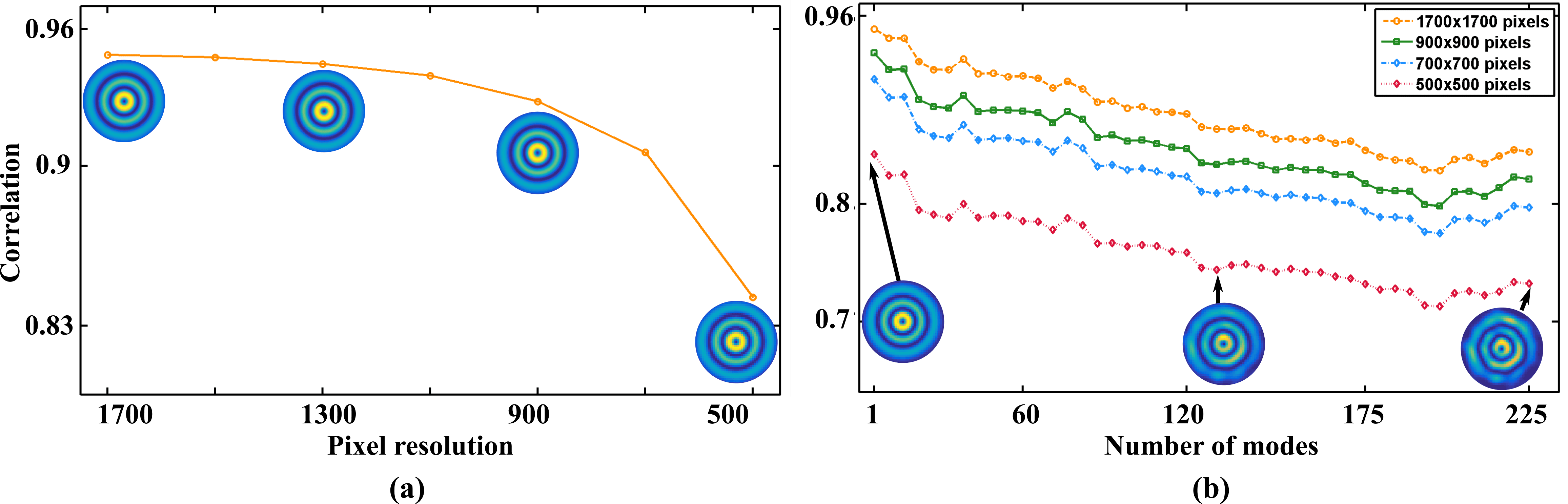}
\caption{\textbf{Correlation as a function of the Resolution.}}
\label{CorPlotSim}
\end{figure}

A simulation of multiplexed beams was carried out using the Fraunhofer approximation of the angular spectrum method (far field), to extend our results to other SLM's screen resolutions. Again, the intensity profile of the simulated mode was correlated against the theoretical. The modes generated through these simulations are, non surprisingly, of higher quality but clearly illustrates a decays in the correlation, as the pixel resolution decreases from $1700$ $\times$ $1700$ to $500$ $\times$ $500$ [Fig. \ref{CorPlotSim}(a)]. 
For this plot, a single mode ($LG_2^1$) was generated and its correlation was computed as the pixel resolution was decreased. The insets shows the intensity profile of this modes at various resolutions. We also simulated, the correlation as a function of the number of multiplexed modes for different resolutions, as shown in Fig. \ref{CorPlotSim}(b). As in the experiment, the correlation coefficient decreases as the number of modes increases. The insets of this figure, shows the intensity profile of the simulated modes for a resolution of $500$ $\times$ $500$ for different number of multiplexed modes.   
\section{Concluding remarks}

Digital holography has emerged as a powerful technique for light beam shaping, prompted by the advent of computer controlled devices such as spatial light modulators. The advantages of light beam shaping through digital devices has been fully exploited in a wide variety of applications\cite{Roadmap,structuredlight,TwPh}. In particular, the simultaneous generation of multiple beams (multiplexing) has enabled applications in fields as optical tweezers and optical communications to mention just a few. In this tutorial we presented a complete description of this technique, while guiding the reader through all the steps involved in the process. First, we introduced two basis sets of spatial modes, Laguerre-Gaussian and Hermite Gaussian, that provided with the modes to exemplify the multiplexing principle. Incidentally, each mode set forms a complete orthogonal solution set of the paraxial wave equations in cylindrical and Cartesian coordinates, respectively. Next, we outlined step by step the process of digital hologram generation to produce theses modes. Moreover, since most optical fields requires the shaping of both, amplitude and phase degrees of freedom for an accurate reproduction, we outlined two commonly used techniques to perform such modulations. The first comprises the generation of phase-only holograms whereas the second one consists in the generation of holograms where the amplitude modulation is embedded in the phase to generate a complex amplitude modulation hologram. Of these two techniques, complex amplitude modulation reproduces optical modes with higher qualities. Not surprising, its implementation is rather challenging. Nevertheless several approaches have being derived over the years to outline its implementation. Here we only focused on one of these methods, described in the text, which produces spatial modes with high quality. Both described techniques allows the multiplexing of many beams and the methodology to accomplish this was also explained to detail in the main text. Moreover, we analyzed the quality of the generated modes while increasing the number of multiplexed modes from 1 up to 225. This analysis provides with useful information of how the mode quality decreases as the number of multiplexed modes increases. Further analysis of the mode quality as function of the resolution of the screen's digital device were carried out through the angular spectrum method. These revealed, as expected, that the quality of the modes decays as the resolution decreases. 

\section*{Acknowledgments}
The authors acknowledge financial support from CONACyT, the Claude Leon foundation, CSIR-DST, the South African National Research Foundation and the Center of Excellence in Strong Materials.

\section*{References}
\providecommand{\newblock}{}


\begin{thebibliography}{10}
\expandafter\ifx\csname url\endcsname\relax
  \def\url#1{{\tt #1}}\fi
\expandafter\ifx\csname urlprefix\endcsname\relax\def\urlprefix{URL }\fi
\providecommand{\eprint}[2][]{\url{#2}}

\bibitem{Dickey2014}
Dickey F~M 2014 {\em Laser Beam Shaping: Theory and Techniques\/} (CRC Press,
  New York)

\bibitem{Forbes}
Forbes A 2014 {\em Laser beam propagation: generation and propagation of
  customized light\/} (CRC Press)

\bibitem{Ngcobo2013}
Ngcobo S, Litvin I, Burger L and Forbes A 2013 {\em Nat. Commun.\/} {\bf 4}
  2289

\bibitem{Forbes2016}
Forbes A, Dudley A and McLaren M 2016 {\em Advances in Optics and Photonics\/}
  {\bf 8} 200--227 ISSN 1943-8206

\bibitem{OptExpress}
Mirhosseini M, Magana-Loaiza O~S, Chen C, Rodenburg B, Malik M and Boyd R 2013
  {\em Opt. Express\/} {\bf 21}

\bibitem{Matsumoto2008}
Matsumoto N, Ando T, Inoue T, Ohtake Y, Fukuchi N and Hara T 2008 {\em J. Opt.
  Am. A\/} {\bf 25} 1642--1651

\bibitem{Ando2009}
Ando T, Ohtake Y, Matsumoto N, Inoue T and Fukuchi N 2009 {\em Opt. Lett.\/}
  {\bf 34} 34--36 \urlprefix\url{http://ol.osa.org/abstract.cfm?URI=ol-34-1-34}

\bibitem{McDonald2013}
McDonald C, McPherson M, McDougall C and McGloin D 2013 {\em Journal of
  Optics\/} {\bf 15} 035708
  \urlprefix\url{http://stacks.iop.org/2040-8986/15/i=3/a=035708}

\bibitem{Grieve2009}
Grieve J~A, Ulcinas A, Subramanian S, Gibson G~M, Padgett M~J, Carberry D~M and
  Miles M~J 2009 {\em Opt. Express\/} {\bf 17} 3595--3602
  \urlprefix\url{http://www.opticsexpress.org/abstract.cfm?URI=oe-17-5-3595}

\bibitem{Bowman2011}
Bowman R~W, Gibson G, Carberry D, Picco L, Miles M and Padgett M~J 2011 {\em
  Journal of Optics\/} {\bf 13} 044002
  \urlprefix\url{http://stacks.iop.org/2040-8986/13/i=4/a=044002}

\bibitem{Thalhammer2013}
Thalhammer G, Bowman R~W, Love G~D, Padgett M~J and Ritsch-Marte M 2013 {\em
  Opt. Express\/} {\bf 21} 1779--1797
  \urlprefix\url{http://www.opticsexpress.org/abstract.cfm?URI=oe-21-2-1779}

\bibitem{Radwell2014}
Radwell N, Brickus D, Clark T~W and Franke-Arnold S 2014 {\em Opt. Express\/}
  {\bf 22} 12845--12852
  \urlprefix\url{http://www.opticsexpress.org/abstract.cfm?URI=oe-22-11-12845}

\bibitem{Curtis2002}
Curtis J~E, Koss B~A and Grier D~G 2002 {\em Optics Communications\/} {\bf 207}
  169 -- 175 ISSN 0030-4018
  \urlprefix\url{http://www.sciencedirect.com/science/article/pii/S0030401802015249}

\bibitem{Grier2002}
Curtis J, Koss B and Grier D 2002 {\em Opt. Commun.\/} {\bf 207} 169--175

\bibitem{Rosales2014OL}
Rosales-Guzm{\'a}n C, Hermosa N, Belmonte A and Torres J~P 2014 {\em Opt.
  Lett.\/} {\bf 18} 5415--5418

\bibitem{Rosales2014OE}
Rosales-Guzm{\'a}n C, Hermosa N, Belmonte A and Torres J~P 2014 {\em Opt.
  Express\/} {\bf 22} 16504--16509

\bibitem{Perez-Vizcaino2012}
P\'{e}rez-Vizca\'{i}no J, Mendoza-Yero O, Mart\'{i}nez-Cuenca R,
  Mart\'{i}nez-Le\'{o}n L, Tajahuerce E and Lancis J 2012 {\em J. Display
  Technol.\/} {\bf 8} 539--545
  \urlprefix\url{http://jdt.osa.org/abstract.cfm?URI=jdt-8-9-539}

\bibitem{Dudley2014}
Dudley A, Milione G, Alfano R~R and Forbes A 2014 {\em Opt. Express\/} {\bf 22}
  14031--14040
  \urlprefix\url{http://www.opticsexpress.org/abstract.cfm?URI=oe-22-11-14031}

\bibitem{Schulze2015}
Schulze C, Roux F~S, Dudley A, Rop R, Duparr\'e M and Forbes A 2015 {\em Phys.
  Rev. A\/} {\bf 91}(4) 043821
  \urlprefix\url{http://link.aps.org/doi/10.1103/PhysRevA.91.043821}

\bibitem{Webster2017}
Webster J, Rosales-Guzm\'{a}n C and Forbes A 2017 {\em Opt. Lett.\/} {\bf 42}
  675--678 \urlprefix\url{http://ol.osa.org/abstract.cfm?URI=ol-42-4-675}

\bibitem{Ndagano2016}
Ndagano B, Sroor H, McLaren M, Rosales-Guzm\'{a}n C and Forbes A 2016 {\em Opt.
  Lett.\/} {\bf 41} 3407--3410
  \urlprefix\url{http://ol.osa.org/abstract.cfm?URI=ol-41-15-3407}

\bibitem{Cox2016}
Cox M~A, Rosales-Guzm\'{a}n C, Lavery M~P~J, Versfeld D~J and Forbes A 2016
  {\em Opt. Express\/} {\bf 24} 18105--18113
  \urlprefix\url{http://www.opticsexpress.org/abstract.cfm?URI=oe-24-16-18105}

\bibitem{Mas2011}
Mas J, Roth M~S, Mart\'{i}n-Badosa E and Montes-Usategui M 2011 {\em Appl.
  Opt.\/} {\bf 50} 1417--1424
  \urlprefix\url{http://ao.osa.org/abstract.cfm?URI=ao-50-10-1417}

\bibitem{Davis1989}
Davis J~A, Cottrell D~M, Lilly R~A and Connely S~W 1989 {\em Opt. Lett.\/} {\bf
  14} 420--422 \urlprefix\url{http://ol.osa.org/abstract.cfm?URI=ol-14-9-420}

\bibitem{Maurer2010}
Maurer C, Khan S, Fassl S, Bernet S and Ritsch-Marte M 2010 {\em Opt.
  Express\/} {\bf 18} 3023--3034
  \urlprefix\url{http://www.opticsexpress.org/abstract.cfm?URI=oe-18-3-3023}

\bibitem{Maurer2011}
Maurer C, Jesacher A, Bernet S and Ritsch-Marte M 2011 {\em Laser \& Photonics
  Reviews\/} {\bf 5} 81--101 ISSN 1863-8899
  \urlprefix\url{http://dx.doi.org/10.1002/lpor.200900047}

\bibitem{Trichili2016}
Trichili A, Rosales-Guzm{\'{a}}n C, Dudley A, Ndagano B, Salem A~B, Zghal M and
  Forbes A 2016 {\em Sci. Rep.\/}  2--7

\bibitem{DGrier2003}
Grier D 2003 {\em Nature\/} {\bf 424} 810--816

\bibitem{Cismar2010}
Čižmár T, Romero L~C~D, Dholakia K and Andrews D~L 2010 {\em Journal of
  Physics B: Atomic, Molecular and Optical Physics\/} {\bf 43} 102001
  \urlprefix\url{http://stacks.iop.org/0953-4075/43/i=10/a=102001}

\bibitem{Eriksen2002}
Eriksen R~L, Daria V~R and Gl\"{u}ckstad J 2002 {\em Opt. Express\/} {\bf 10}
  597--602
  \urlprefix\url{http://www.opticsexpress.org/abstract.cfm?URI=oe-10-14-597}

\bibitem{Wang2}
Wang J, Li S, Luo M, Liu J, Zhu L, Li C, Xie D, Yang Q, Yu S, Sun J, Zhang X,
  Shieh W and Willner A 2014 {N}-dimentional multiplexing link with
  1.036-{P}bit/s transmission capacity and 112.6-bit/s/{H}z spectral efficiency
  using {OFDM}-8{QAM} signals over 368 {WDM} pol-muxed 26 {OAM} modes {\em
  Optical Communication (ECOC), 2014 European Conference on\/} pp 1--3

\bibitem{Siegman}
Siegman A~E 1986 {\em {Lasers}\/} (Standfor Iniversity)

\bibitem{Ohtake2007}
Ohtake Y, Ando T, Fukuchi N, Matsumoto N, Ito H and Hara T 2007 {\em Opt.
  Lett.\/} {\bf 32} 1411--1413
  \urlprefix\url{http://ol.osa.org/abstract.cfm?URI=ol-32-11-1411}

\bibitem{Arrizon}
Arriz\'{o}n V, Ruiz U, Carrada R and Gonz\'{a}lez L~A 2007 {\em J. Opt. Soc.
  Am. A\/} {\bf 24} 3500--3507

\bibitem{Davis2003}
Davis J~A, Valad\'{e}z K~O and Cottrell D~M 2003 {\em Appl. Opt.\/} {\bf 42}
  2003--2008 \urlprefix\url{http://ao.osa.org/abstract.cfm?URI=ao-42-11-2003}

\bibitem{Putten2008}
van Putten E~G, Vellekoop I~M and Mosk A~P 2008 {\em Appl. Opt.\/} {\bf 47}
  2076--2081 \urlprefix\url{http://ao.osa.org/abstract.cfm?URI=ao-47-12-2076}

\bibitem{Clark2016}
Clark T~W, Offer R~F, Franke-Arnold S, Arnold A~S and Radwell N 2016 {\em Opt.
  Express\/} {\bf 24} 6249--6264
  \urlprefix\url{http://www.opticsexpress.org/abstract.cfm?URI=oe-24-6-6249}

\bibitem{Aguirre2015}
Aguirre-Olivas D, {n}or G~M~V, de-la Llave D~S and Arriz\'{o}n V 2015 {\em
  Appl. Opt.\/} {\bf 54} 8444--8452
  \urlprefix\url{http://ao.osa.org/abstract.cfm?URI=ao-54-28-8444}

\bibitem{Roadmap}
Rubinsztein-Dunlop H, Forbes A, Berry M~V, Dennis M~R, Andrews D~L, Mansuripur
  M, Denz C, Alpmann C, Banzer P, Bauer T, Karimi E, Marrucci L, Padgett M,
  Ritsch-Marte M, Litchinitser N~M, Bigelow N~P, Rosales-Guzm{\'a}n C, Belmonte
  A, Torres J~P, Neely T~W, Baker M, Gordon R, Stilgoe A~B, Romero J, White
  A~G, Fickler R, Willner A~E, Xie G, McMorran B and Weiner A~M 2017 {\em
  Journal of Optics\/} {\bf 19} 013001
  \urlprefix\url{http://stacks.iop.org/2040-8986/19/i=1/a=013001}

\bibitem{structuredlight}
Andrews D~L 2008 {\em {Structured light and its applications}\/} (Academic
  Press-Elsevier, Burlington.)

\bibitem{TwPh}
Torres J~P and Torner L 2011 {\em {Twisted Photons}\/} (Wiley-VCH, Bristol.)

\end{thebibliography}
\end{document}